\documentclass[aps,twocolumn]{revtex4}

\usepackage{graphicx}
\usepackage{color}
\begin{document}

\bibliographystyle{apsrev}

\title{Discriminating between a Stochastic Gravitational Wave Background \\ and Instrument Noise}
\author{\surname {Matthew} R. Adams and {Neil} J. Cornish}

\affiliation{Department of Physics, Montana State University, Bozeman, MT 59717}
\date{\today}

\begin{abstract}

The detection of a stochastic background of gravitational waves could significantly impact
our understanding of the physical processes that shaped the early Universe. The challenge lies
in separating the cosmological signal from other stochastic processes such as instrument
noise and astrophysical foregrounds. One approach is to build two or more detectors and
cross correlate their output, thereby enhancing the common gravitational wave signal
relative to the uncorrelated instrument noise. When only one detector is available,
as will likely be the case with the Laser Interferometer Space Antenna (LISA), alternative
analysis techniques must be developed. Here we show that models of the noise and signal transfer
functions can be used to tease apart the gravitational and instrument noise contributions. We
discuss the role of gravitational wave insensitive ``null channels'' formed from particular
combinations of the time delay interferometry, and derive a new combination that maintains
this insensitivity for unequal arm length detectors. We show that, in the absence of astrophysical
foregrounds, LISA  could detect signals with energy densities as low as
$\Omega_{\rm gw} = 6 \times 10^{-13}$ with just one month of data.  We describe an end-to-end Bayesian
analysis pipeline that is able to search for, characterize and assign confidence levels for the detection
of a stochastic gravitational wave background, and demonstrate the effectiveness of this
approach using simulated
data from the third round of Mock LISA Data Challenges.
\end{abstract}

\maketitle

\section{Introduction}
One of the most exciting prospects for the Laser Interferometer Space Antenna (LISA)
observatory is the possibility of detecting a stochastic gravitational wave background from
the early Universe. Just as studies of the cosmic microwave background have revolutionized our
understanding of cosmology \cite{Komatsu:2008hk}, the detection of a gravitational wave background
would provide unique insight into the processes that shaped the early universe~\cite{Hogan:1998dv,Maggiore:1999vm,Cutler:2002me,Hughes:2002yy,Buonanno:2003th,Chongchitnan:2006pe}.

The challenge lies in separating a stochastic background from instrument noise and foreground sources.  When multiple independent detectors are available, as is the case with the ground based interferometers, the signal from one interferometer can be used as a (noise corrupted) template for a second interferometer.  The common gravitational wave signal will combine coherently, while the contributions from instrument noise will average to
zero~\cite{Christensen:1992wi,Flanagan:1993ix,Cornish:2001qi}

With a single detector, such as LISA, the cross correlation technique will not be available. The key to detecting a stochastic background with LISA is the development of a detailed understanding of how instrument noise and gravitational wave signals manifest in the various interferometry channels.  Here we show how a model for the six
cross-spectra of the Time Delay Interferometry (TDI) \cite{Tinto:1999yr,Estabrook:2000ef} channels can be
used to tease apart a stochastic background from instrument noise.  Key to this approach are the very different
geometrical transfer functions that affect the instrument noise and a stochastic gravitational wave background.
While the instrument noise spectra could in principle conspire to mimic the geometrical transfer functions
of the signal, or vice-versa, this is unlikely to occur in practice as the signal and noise models have
very distinct and complicated transfer functions. The robustness of our approach can be enhanced by
developing informative priors for the noise and signal spectra. This can be done
through a combination of preflight testing, on-orbit commissioning studies, and theoretical modeling.
We use the TDI cross-spectra to compute a likelihood function, which is multiplied by the priors to
yield the posterior distributions for the parameters describing the instrument noise and cosmological
background. 

Models for the TDI cross-spectra have previously been considered in the context of LISA instrument
noise determination~\cite{Sylvestre:2003in}, and the insensitivity of certain TDI variables to gravitational
wave signals have been put forward as a technique for discriminating between a stochastic
background and instrument  noise~\cite{Tinto:1999yr,Estabrook:2000ef,Sylvestre:2003in}.
Our approach extends the noise characterization study of Sylvestre and Tinto \cite{Sylvestre:2003in}
to include the signal cross-spectra, and improves upon the simple estimator used by Hogan and
Bender~\cite{Hogan:2001jn} by using an optimal combination of all six cross-spectra.
Our approach is able to detect stochastic signals buried well below the instrument noise.

In the remainder of the paper we describe our noise and signal models for both equal arm and unequal arm
LISA interferometers, and develop an end-to-end Bayesian analysis pipeline that is able to search for,
characterize and assign confidence levels for the detection of a stochastic gravitational wave background.
We apply our approach to simulated data from the third round of Mock LISA Data Challenge, and show that
we are able to accurately recover the stochastic signal and independently measure the position and
acceleration noise levels in each arm of the interferometer.  We carry out our analysis using a noise
orthogonal $A,E,T$ set of TDI channels.  In this basis, and for equal arm-lengths, the $T$ channel is insensitive
to gravitational wave signals at low frequencies. For the first time, we show that a modified
signal insensitive $T$ channel can be found for unequal arm-lengths.

We close with a discussion of how our approach can be extended to account for the galactic foreground.
While several tens of thousands of galactic sources can be individually resolved and regressed from the data,
the remainder will form a confusion noise for LISA~\cite{Crowder:2006eu}.  We discuss how the anisotropy of
this foreground signal, along with prior information about the spectral amplitude provided by the resolved
systems, can be used to separate the galactic foreground from an isotropic background, but defer a
detailed analysis to a future publication.

\section{Equal Arm LISA}
\subsection{Noise Model}
The LISA constellation will be composed of three satellites in an approximately equilateral triangle configuration.  Laser beams will be transmitted between each pair of satellites, and interferometry signals will be formed using these beams.  Each of the three satellites will have two proof masses, one for each of the two
incoming laser beams.  We are looking to detect variations in the light travel time, or equivalently, the
distance, between the proof masses along each arm of the interferometer.
Noise enters the measurement when the proof masses move in response to local disturbances, and in the process
of measuring the phase of the laser light.  The various LISA noise sources are discussed in several
references~\cite{Bender:1998, Danzmann:2003, Purdue:2007zz}.

As is commonly done, we group all the noise sources into two categories, position and acceleration.  Each proof mass will have a position and an acceleration noise
associated with it, making a total of six position and six acceleration noise levels.  These twelve noise levels will be the parameters in our noise model.  We assume that we understand the noise spectra of each of the instrument components, through some combination of pre-launch testing, on-board instrument characterization and theoretical modeling.  In reality it may be necessary to include additional parameters in the noise model to account for uncertainties in the spectral shape of the individual noise spectra.  However, if the priors on these additional parameters are narrowly peaked, there will be little impact on our ability to detect a stochastic background signal.

For this study we adopted the model used for the Mock LISA Data Challenges (MLDC) \cite{:2008sn}.  The position
noise affecting each proof mass is assumed to be white, with a nominal spectral density of
\begin{equation}
S_p(f) = 4 \times 10^{-42} \mbox{Hz}^{-1}.
\end{equation}
The acceleration noise is taken as white above 0.1 mHz, with a red component below this
frequency. Integrated to give an effective position noise, the proof mass disturbances on
each test mass have a nominal spectral density of
\begin{equation}
S_a(f) = 9 \times 10^{-50} \left(1 + \left(\frac{10^{-4} {\rm Hz}}{f}\right)^2\right)
\left(\frac{\mbox{mHz}}{2 \pi f}\right)^4 \mbox{Hz}^{-1}
\end{equation}
The precise level of each contribution is to be determined from the data. A more realistic model for
the noise contributions would include a parameterized model for the frequency dependence, with the
model parameters to be inferred from the data. Allowing for this additional freedom would weaken the
bounds that can be placed on the contribution from the stochastic background.

We derive here the transfer functions describing how these twelve noise levels enter the data stream. 
We start by writing down the phase output $\Phi_{ij}(t)$, for the link connecting spacecraft $i$ and $j$:
\begin{eqnarray}
\Phi_{ij}(t) &=& C_i(t-L_{ij})-C_j(t) + \psi_{ij}(t) \\
             &+& n^p_{ij}-\hat{x}_{ij}\cdot(\vec{n}^a_{ij}(t)-\vec{n}^a_{ij}(t-L_{ij}))
\end{eqnarray}
Here the $C_i$ are the laser phase noises, $\psi_{ij}$ is the gravitational wave strain, and $n^p_{ij}$
and $\vec{n}^a_{ij}$ denote the position and acceleration noise.  The laser 
phase noise is canceled by using the Time Delay Interferometry(TDI) variables developed by Armstrong,
Estabrook, and Tinto\cite{Tinto:1999yr,Estabrook:2000ef}.  The gravitational wave strain is assumed to
be uncorrelated with the noise and will have an expectation value of zero when multiplied by anything
other than itself.  

A Michelson signal can be formed at any of the three vertices by combining the phase at that detector with
the time delayed signal from the two detectors at the ends of the two adjacent arms:
\begin{equation}
M_1(t) = \Phi_{12}(t-L_{12}) + \Phi_{21}(t) - \Phi_{12}(t-L_{13}) - \Phi_{31}(t)
\end{equation}
The TDI channels which cancel laser phase noise for an equal arm LISA ($L_{ij}=L=5 \cdot 10^6 \mbox{ km}$)
are formed by subtracting a time delayed Michelson signal as follows:
\begin{equation}
X(t) = M_1(t) - M_1(t-2L)
\end{equation}
Moving to the frequency domain, the signal at vertex 1 can be written as
\begin{eqnarray}
X(f) &=& 2i\sin\left(\frac{f}{f_*}\right)e^{f/f_*}\left[e^{f/f_*}(n^p_{13}-n^p_{12})+n^p_{31}-n^p_{21}\right]   \nonumber \\
  & & +4i\sin\left(\frac{2f}{f_*}\right)e^{2f/f_*}\Bigg[(n^a_{12}+n^a_{13}) \nonumber \\
	& & -(n^a_{21}+n^a_{31})\cos\left(\frac{f}{f_*}\right)\Bigg]
\end{eqnarray}
Here $f/f_* = c/(2\pi L)$. The other TDI channels, $Y$ and $Z$, are given by permuting indices
in the expression for the X channel. Alternative combinations of the phase measurements can be
used to derive noise orthogonal TDI variables \cite{Prince:2002hp}. Here we use a different set of
noise orthogonal channels formed from combinations of the $X$,$Y$, and $Z$ channels:
\begin{eqnarray}
\label{AET}
A &=& \frac{1}{3} (2X-Y-Z) \nonumber \\
E &=& \frac{1}{\sqrt{3}} (Z-Y) \nonumber \\
T &=& \frac{1}{3} (X+Y+Z)
\end{eqnarray}
We calculate the six cross-spectral densities for these channels in the appendix. As an example,
we quote here the position and acceleration noise contributions to $\langle A A^* \rangle$: 
\begin{eqnarray}
 \left<AA^*_p\right> &=& \frac{4}{9} \sin^2{\left(\frac{f}{f_*}\right)}                             \nonumber \\
        & & \Bigg\{  \cos{\left(\frac{f}{f_*}\right)}\big[4(S^p_{21}+S^p_{12}+S^p_{13}+S^p_{31}) \nonumber\\
        & &-2(S^p_{23}+S^p_{32})\big]  \nonumber \\
        & &   +5(S^p_{21}+S^p_{12}+S^p_{13}+S^p_{31}) \nonumber \\
        & & +2(S^p_{23}+S^p_{32})\Bigg\}
\end{eqnarray}
and		
\begin{eqnarray}
\left<AA^*_a\right> &=& \frac{16}{9}\sin^2{\left(\frac{f}{f_*}\right)} \nonumber \\ 
       & & \bigg\{(\cos\left(\frac{f}{f_*}\right)\bigg[4(S^a_{12}+S^a_{13}+S^a_{31}+S^a_{21}) \nonumber \\ 
       & & -2(S^a_{23}+S^a_{32})\bigg] \nonumber \\       
       & & +\cos\left(\frac{f}{f_*}\right)\bigg[\frac{3}{2}(S^a_{12}+S^a_{13}+S^a_{23}+S^a_{32}) \nonumber \\ 
       & & +2(S^a_{31}+S^a_{21})\bigg] \nonumber \\   
       & & +\frac{9}{2}(S^a_{12}+S^a_{13})+3(S^a_{31}+S^a_{21}) \nonumber \\ 
			 & & +\frac{3}{2}(S^a_{23}+S^a_{32})\bigg\}
\end{eqnarray}
where $S_{ij}(f)=\langle n_{ij}(f)n_{ij}^*(f)\rangle$. Our model for the noise
depends on the twelve noise spectral density levels, which we assume fully describe the
LISA instrument noise. Note that the sums of the noise contributions
in each arm of the interferometer, such as $(S^a_{23}+S^a_{32})$, that are strongly
constrained. The differences are very weakly constrained by the model. Figure 1 compares
our model spectra to simulated LISA noise spectra from the
MLDC training data.
\begin{figure}[htbp]
   \centering
   \includegraphics[width=2.3in,angle=270] {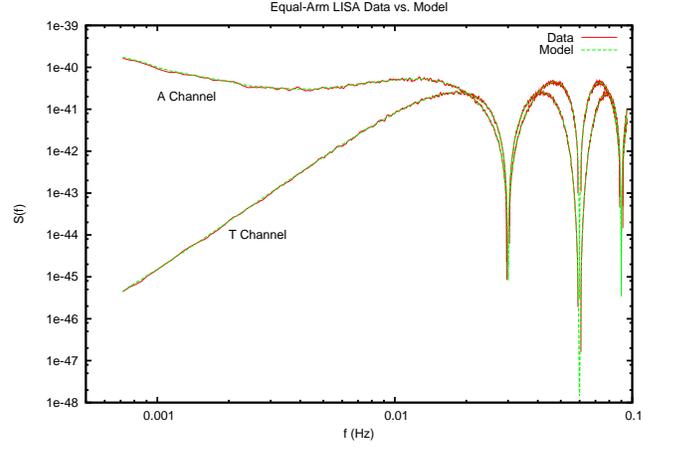} 
   \caption{Our model for the noise in the $A$ and $T$ channels compared to smoothed spectra formed
from the MLDC training data.}
\end{figure}

\subsection{Stochastic Background Model}
In this section we re-derive the LISA response to a stochastic gravitational wave background. We later extend
this calculation to take into account the effect of unequal arm-lengths.
We begin by expanding the gravitational wave background in plane waves:
\begin{equation}
\label{hstrain}
h_{ij}(t,\vec{x}) = \sum_P \int_{-\infty}^{\infty} df 
  \int d\Omega \tilde{h}(f,\hat{\Omega})e^{-2 \pi f(t-\hat{\Omega}\cdot \vec{x})}\epsilon_{ij}^P(\hat{\Omega})
\end{equation}       
Here the $\epsilon_{ij}$ are the components of the polarization tensor and $P$ sums over the two
polarizations.  The polarization tensors are formed by using the basis vectors $\hat{u}$
and $\hat{v}$ and the sky location vector $\hat{\Omega}$,
\begin{eqnarray}
\hat{u} &=& \cos{\theta}\cos{\phi}\hat{x} + \cos{\theta}\sin{\phi}\hat{y}-\sin{\theta}\hat{z} \nonumber \\
\hat{v} &=& \sin{\phi} \hat{x} - \cos{\phi} \hat{y} \nonumber \\
\hat{\Omega} &=& \sin{\theta}\cos{\phi}\hat{x} + \sin{\theta}\sin{\phi}\hat{y} + \cos{\theta} \hat{z}
\end{eqnarray}
The polarization tensors are formed as
\begin{equation}                               
\epsilon^+(\hat{\Omega},\psi) = e^+(\hat{\Omega})\cos(2\psi) - e^{\times}(\hat{\Omega})\sin(2\psi)
\end{equation}	
and
\begin{equation}
\epsilon^{\times}(\hat{\Omega},\psi) = e^+(\hat{\Omega})\sin(2\psi) + e^{\times}(\hat{\Omega})\cos(2\psi)
\end{equation}
where $\psi$ is the polarization angle and $e^+$ and $e^{\times}$ are given by:
\begin{equation}
e^+ = \hat{u} \otimes \hat{u} - \hat{v} \otimes \hat{v}
\end{equation}
and
\begin{equation}
e^{\times} = \hat{u} \otimes \hat{v} + \hat{v} \otimes \hat{u} \, .
\end{equation}

The response registered in an interferometry channel can be written as
\begin{equation}
S_i(t) = {\bf D}_i(\hat{\Omega},f):{\bf h}(f,\vec{x})
\end{equation}
where
\begin{eqnarray}
{\bf D}_i(\hat{\Omega},f) &=& \frac{1}{2}(\hat{r}_{ij}\otimes\hat{r}_{ij}) \mathcal{T}(\hat{r}_{ij}\cdot\hat{k},f)  \nonumber \\
                          &-& \frac{1}{2}(\hat{r}_{il}\otimes\hat{r}_{il}) \mathcal{T}(\hat{r}_{il}\cdot\hat{k},f)  \nonumber
\end{eqnarray}
and
\begin{eqnarray}
\mathcal{T} &=& \frac{1}{2} \mbox{sinc}(\frac{\omega}{2\omega_{ij}}(1-\hat{\Omega}\cdot \hat{r}_{ij}(t_i))\exp{i\frac{\omega}{2\omega_{ij}}(3+\hat{\Omega}\cdot\hat{r}_{ij})}   \nonumber  \\
    &+& \frac{1}{2} \mbox{sinc}(\frac{\omega}{2\omega_{ij}}(1+\hat{\Omega}\cdot \hat{r}_{ij}(t_i))\exp{i\frac{\omega}{2\omega_{ij}}(1+\hat{\Omega}\cdot\hat{r}_{ij})})\, . \nonumber
\end{eqnarray}
 
The signal cross spectra are given by 
\begin{equation}
\left<S_i(f),S_j(f)\right> = S_h(f)R_{ij}(f)
\end{equation}
where
\begin{equation} \label{R}
R_{ij}(f) = \sum_P \int \frac{d\Omega}{4\pi} F^P_i(\hat{\Omega},f) F^{P^*}_j(\hat{\Omega},f)
\end{equation}
and
\begin{equation}
\label{Sh}
S_h(f) = \frac{3H_0^2}{4\pi^2}\frac{\Omega_{gw}(f)}{f^3} \, .
\end{equation}
Here $\Omega_{gw}(f)$ is the energy density in gravitational waves per logarithmic frequency interval,
scaled by the closure density. We assumed a simple power law
behavior: $\Omega_{gw}(f) =  (f/1{\rm mHz})^n \, \Omega_{gw}$,
where $\Omega_{gw}$ denotes the level at $1 {\rm mHz}$ and $n$ is the spectral index.

The beam pattern functions are given by
\begin{equation}	
F^P_i(\hat{\Omega},f) = \mathbf{D}_i(\hat{\Omega},f):\mathbf{e}^P(\hat{\Omega}) \, .
\end{equation}
In general, the integral in (\ref{R}) must be performed numerically,
though we can develop analytic expressions in the low frequency limit: 
\begin{eqnarray}
R_{AA} = R_{EE} &=& 4\sin^2\left(\frac{f}{f_*}\right)\bigg[\frac{3}{10}-\frac{169}{1680}\left(\frac{f}{f_*}\right)^2 \\ \nonumber
                & & +\frac{85}{6048}\left(\frac{f}{f_*}\right)^4 - \frac{178273}{159667200}\left(\frac{f}{f_*}\right)^6          \\ \nonumber
                & & + \frac{19121}{24766560000}\left(\frac{f}{f_*}\right)^8 + ...\bigg]
\end{eqnarray}
and
\begin{eqnarray}
R_{TT} &=&  4\sin^2\left(\frac{f}{f_*}\right)\bigg[\frac{1}{12096}\left(\frac{f}{f_*}\right)^6  \\ \nonumber
       & & -\frac{61}{4354560}\left(\frac{f}{f_*}\right)^8 + ...\bigg]
\end{eqnarray}
The $A$,$E$, and $T$ channels were created to be noise orthogonal, but they also happen to be
signal orthogonal in the equal arm case.  All of the cross terms in the response function
$R_{AE}$, $R_{AT}$ etc. are zero in the equal arm-length limit.
Sensitivity curves for the various channels are generated by plotting
\begin{equation}
h_{K} = \sqrt{\frac{S_{KK}(f)}{R_{KK}(f)}}
\end{equation}
where $S_{KK}$ and $R_{KK}$ are the noise and signal spectral densities in the $K$ channel.
Figure 2 shows the sensitivity curves for the $A, E$ and $T$ channels along with a scale
invariant gravitational wave background with $\Omega_{\rm gw} = 10^{-10}$.
We see that $T$ channel is insensitive to the gravitational wave background for $f<f_*$.
\begin{figure}[htbp]
\label{EqSens}
   \centering
   \includegraphics[width=2.3in,angle=270] {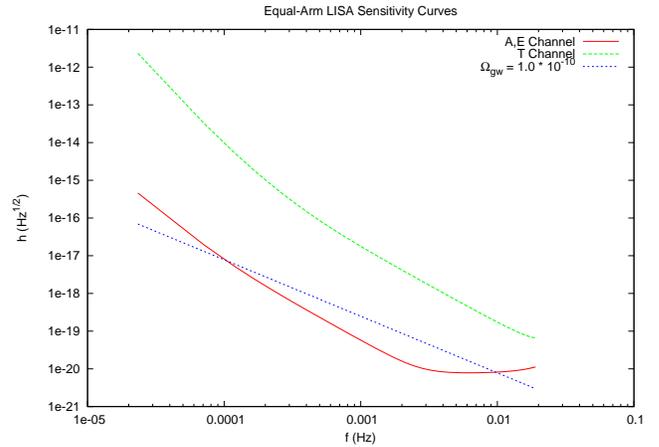} 
   \caption{The sensitivity curve for the A, E and T channels, showing the insensitivity of the
T channel to a gravitational wave signal.}
\end{figure}

\section{Search and Characterization}

For Gaussian signals and noise, the likelihood of measuring cross-spectra $<X_i X_j>$ is given by:
\begin{equation}
p({\bf X} \vert \vec{x}) = \Pi \frac{1}{(2\pi)^{N/2}|C|}\exp\left(X_i C^{-1}_{ij} X_j\right)
\end{equation}
where $C$ is the noise correlation matrix, $X_i = \{A,E,T\}$, $N$ is the number of samples in each
channel and $\vec{x}\rightarrow (S^a_i,S^p_j,\Omega_{gw},n)$ denotes the parameters in our model.
The noise correlation matrix is given by:
\begin{equation}
C_{ij} = \left ( \begin{array}{lcr}
									\langle AA\rangle & \left<AE\right> & \left<AT\right> \\
									\left<EA\right> & \left<EE\right> & \left<ET\right> \\
									\left<TA\right> & \left<TE\right> & \left<TT\right> \\
           \end{array} \right )\, .
\end{equation} 
Combined with a prior for the model parameters, $p(\vec{x})$, we are able to generate samples from the
posterior distributions function $p(\vec{x} \vert {\bf X}) = p({\bf X} \vert \vec{x}) p(\vec{x})/p({\bf X})$
using a Parallel Tempered Markov Chain Monte Carlo algorithm~\cite{Littenberg:2009bm}. Lacking a detailed
instrument model or relevant experimental data, we choose to use uniform priors for the 12 instrument
noise levels, allowing a factor of ten variation above and below the nominal levels. The energy density was
taken to be uniform in $\ln(\Omega_{gw})$ across the range $[-30,-24.5]$. We considered two models for
the spectral slope, either assuming a scale invariant background and fixing $n=0$, or allowing
the spectral index $n$ to be uniform in the range $[-1,1]$.

For the proposal distribution we used a mixture of uniform draws from the full prior range, and draws
from a multivariate Gaussian distribution computed from the Fisher Information
Matrix~\cite{Littenberg:2009bm}.
Correlations between the parameters, and the frequency dependence of the spectra complicate the
computation of the Fisher Matrix, but the basic idea can be understood by
considering zero mean white
noise with variance $\sigma$. The relevant question is, how well can the noise level $\sigma^2$ be
determined from $N$ noise samples? The likelihood of observing the data $\{ x_i \}$ is
\begin{equation}
\mathcal{L} = \frac{1}{(2\pi)^{N/2}\sigma^N}\exp\left( -\frac{\sum_{i=1}^N x_i^2}{2\sigma^2} \right)\, ,
\end{equation}
which yields a maximum likelihood estimate for the noise level of $\hat{\sigma}^2 = \sum_i x_i^2/N$. The
Fisher Matrix has a single element:
\begin{equation}
\Gamma_{\sigma \sigma } = -\frac{ \partial^2 \ln\mathcal{L}}{\partial \sigma^2}\vert_{\rm ML}
= \frac{N}{2 \hat{\sigma}^2} \, .
\end{equation}
Thus, the error in the estimated noise level is $\Delta \sigma^2 = 2 \sigma \Delta \sigma = \sigma^2/\sqrt{N/2}$.
We see that the fractional error in the noise level estimate scales as the square root of
the number of data points. The same is true for the more complicated colored spectra in our LISA
noise model. Noting that the acceleration noise dominates below $\sim 1 {\rm mHz}$, while the
position noise dominates above $\sim 3 {\rm mHz}$, it follows that the effective number of samples
available to constrain the acceleration noise is of order $N_a  \sim 1 {\rm mHz} \times T_{\rm obs}$.
The position noise is far better constrained, with order $N_p  \sim f_{N} \times T_{\rm obs}$ samples,
where $f_{N}$ is the Nyquist frequency of the data. As mentioned earlier, only the sum of the instrument
noise levels in each arm are strongly constrained, so the Fisher matrix approach leads to very large jumps
being proposed in the noise level differences in each arm. To maintain a good acceptance rate, we capped the
variance in the weakly constrained directions to be ten times the variance in the well constrained directions.

\section{Mock LISA Data Challenge}
We are able to test our analysis technique on simulated data from the third round of the Mock LISA Data Challenge.
Specifically, Challenge 3.5 provides month long training and blind data sets of $2^{20}$ samples with 2 second
sampling.  An isotropic gravitational wave background was injected with a level that is approximately 10 times
the nominal noise levels at 1mHz, corresponding to $\Omega_{GW} = 8.95 \times 10^{-12} - 1.66 \times 10^{-11}$
for a Hubble constant of $H_0=70 {\rm km}/{\rm s}/{\rm Mpc}$. The noise levels were drawn from a range
within $\pm 20\%$ of the
nominal values. Note that we used prior ranges far wider range than this as we wanted to test our approach in
a more realistic setting.

\begin{figure}[htbp]
\label{positionN}
   \centering
   \includegraphics[width=2.3in,angle=270] {PositionNoise.epsi} 
   \caption{Histograms showing the posterior distribution functions for the position noise levels,
scaled by the nominal level. One the left are the sums  along each arm, and on the right are
the differences. The vertical lines denote the injected values.}
\end{figure}

\begin{figure}[htbp]
\label{accelerationN}
   \centering
   \includegraphics[width=2.3in,angle=270] {AccelerationNoise.epsi} 
   \caption{Histograms showing the posterior distribution functions for the position noise levels,
scaled by the nominal level. One the left are the sums along each arm, and on the right are
the differences. The vertical lines denote the injected values.}
\end{figure}

\begin{figure}[htbp]
\label{Background}
   \centering
   \includegraphics[width=2.3in,angle=270] {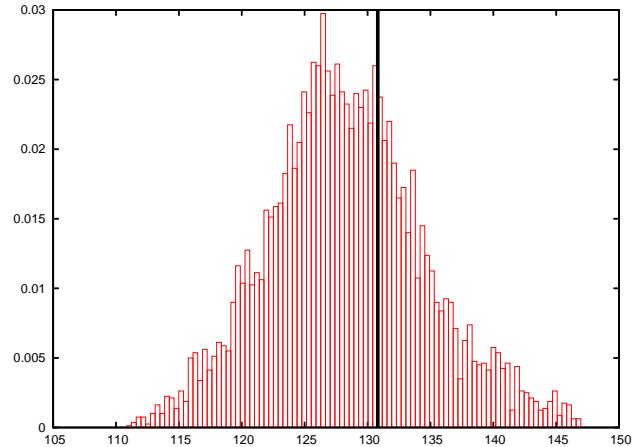} 
   \caption{The posterior distribution for the gravitational wave background level (scaled up by
$10^{13}$). The vertical line denotes the injected values.}
\end{figure}

\begin{figure}[htbp]
\label{Slope}
   \centering
   \includegraphics[width=2.3in,angle=270] {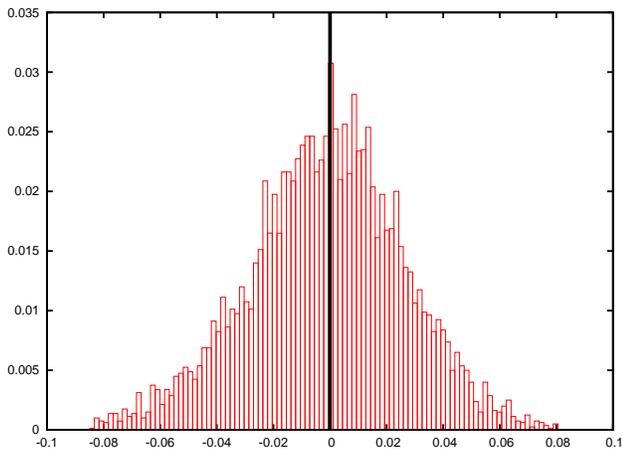} 
   \caption{The posterior distribution for the gravitational wave background slope.}
\end{figure}

\subsection{Training Data Results}
We found that the signal transfer functions in the training data did not match our analytic model. We traced
the problem to the time domain filters that were used to generate the data sets, which introduced additional
transfer functions in the frequency domain. Since the analytic form for these transfer functions have not
been published, we used the training data to estimate the transfer functions and update our signal model.

We used the Bayesian analysis algorithm described above and recovered the PDF for the noise parameters and
stochastic background energy density show in Figures 3-6.  As mentioned earlier,
only the sums of the noise contributions in each arm are constrained (the acceleration and position noise
contributions can be separated though as they have very different transfer functions). Figure 3
provides an example of this by showing that the sums $S^p_{ji}+ S^p_{ij}$ are well constrained, while the
differences $S^p_{ji} - S^p_{ij}$ are poorly constrained. Also note that the position noise levels are far better
determined than the acceleration noise levels, as expected from our Fisher Matrix analysis.

\begin{figure}[htbp]
\label{errorbars}
   \centering
   \includegraphics[width=2.3in,angle=270] {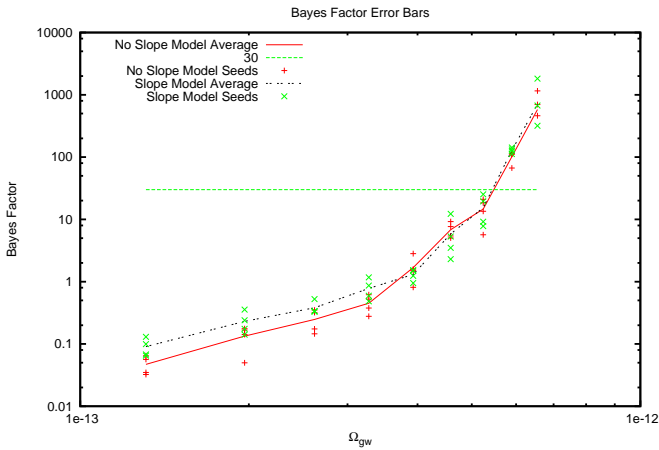} 
   \caption{Bayes factor showing the detectability versus background level.  The line shows the average values of the scattered points.}
   \label{fig:example}
\end{figure}

\subsection{Detection Limits}
Having established that our algorithm can faithfully recover a stochastic background level of
$\Omega_{gw}\sim 10^{-11}$ with one month of data, we next turned to the problem of determining the LISA detection
limit for an approximately scale invariant stochastic gravitational wave background. To do this
we generated a new set of simulated data sets by re-scaling the gravitational wave contribution to the
MLDC training data, and used Bayesian model selection to compared the evidence for two models, $M_0$ -
the data is described by instrument noise alone, and $M_1$ - the data is described by instrument noise
and a stochastic gravitational wave background. Following the approach described in Ref.~\cite{Littenberg:2009bm},
we computed the evidence for each model using thermodynamic integration~\cite{thermo}.
Plots of the Bayes factor, or evidence ratio, as a function of $\Omega_{gw}$ are shown in Figure 7. 
We performed multiple runs with different random number seeds as a way to estimate the numerical
error in our Bayes factors. Our detection confidence becomes very strong (a Bayes factor of 30) for a
background level of $\Omega_{\rm gw} = 6 \times 10^{-13}$ with one month of data. With one year of data the
limit improves to $\Omega_{\rm gw} = 1.7 \times 10^{-13}$.

It is interesting to note that the detection limit does not change if we include the spectral slope
of the background as a model parameter. At first this seems a little surprising, as the simulated data
has a spectral slope of $n=0$, so we would expect the simpler model with $n=0$ to be favored over the
more complicated model with $n$ as a free parameter. Further investigation revealed that the more
complicated model was able to provide a slightly better fit to the data, and that this was enough to
compensate for the additional complexity of the model in the calculation of the model evidence.
Our hypothesis is that our signal model is imperfect because of the need to use numerical fits to the
transfer functions introduce by the simulation software, and that the freedom to adjust the spectral
slope of the background is able to compensate for this imperfection.

\subsection{The role of null channels and 4-link operation}

Previous analyses of stochastic background detection with LISA have emphasized the importance of the
gravitational wave insensitive null channel that can be formed from the Sagnac or Michelson interferometry
channels (to be precise these channels are only null in the zero frequency limit, at non-zero frequencies they
do respond weakly to gravitational signals because of finite arm-length effects). Hogan and Bender showed
how this null channel could be used to construct a statistic that
measures the amplitude of the stochastic background. The importance of null channels has also
been emphasized in the context of searches for un-modeled gravitational wave signals in ground and space
based detectors. For example, in the LIGO-Virgo searches for un-modeled gravitational wave bursts, it has
been shown that the sensitivity can be improved by using the sums and differences of the output from
the two $(H2)$ and four $(H1)$ kilometer detectors at the Hanford site. The null channel $H_- = H1-H2$ is
insensitive to gravitational waves, and has proven useful as a tool to distinguish between instrumental artifacts and
gravitational wave signals~\cite{Abbott:2009zi,Chatterji:2006nh}.

It may therefore seem a little surprising that the null ``$T$'' channel plays no privileged role in
the present analysis. The reason is simple: when using a Gaussian likelihood function the coordinate
transformation in signal space that produces the null channel leaves the likelihood unchanged. 
For example, when we repeat our analysis using the cross spectra for the $\{X,Y,Z\}$ channels we
get results that are {\em identical} to what we found with the $\{A,E,T\}$ channels. It is only when
the instrument noise is not well understood, and there are significant departures from stationarity and
Gaussianity that null channels become important. It would be naive to assume that the LISA data will be
perfectly stationary and Gaussian, and we expect the $T$-channel will play a key role in detector
characterization studies. That is our motivation for the calculation described in the next section,
where we derive a new version of the $T$-channel that is insensitive to gravitational waves for
un-equal arm-lengths.

\begin{figure}[htbp]
\label{BayesXYZ}
   \centering
   \includegraphics[width=2.3in,angle=270] {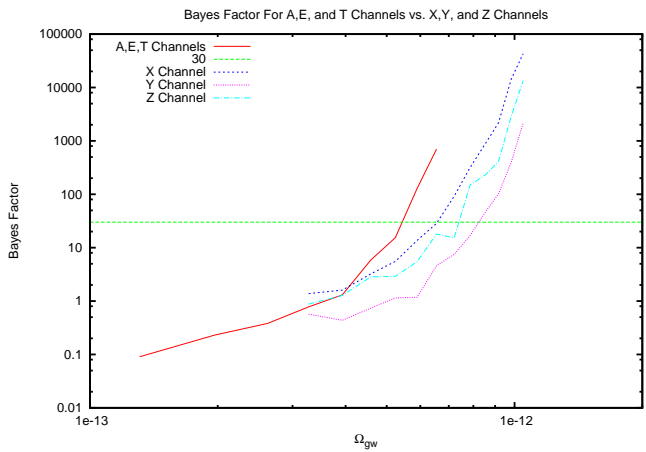} 
   \caption{Bayes factors for the X, Y, and Z channels}
\end{figure}

With the current baseline design for LISA, the failure of one proof mass leaves us with a single
interferometry channel. In most instances, it is then possible to configure the array to produce a
single $X$ type TDI channel (alternatively we can work with a single Beacon, Monitor or Relay channel).
With one channel the number of cross-spectra drops from 6 to 1, and the null direction in signal
space (the $T$ channel) is lost. On the other hand, the number of noise variables drops from
12 to 8 (the effective dimension of the noise model drops from 6 to 4 since only the sums of noise
contributions along each arm can be inferred from the gravitational wave data). To study these competing
effects we repeated our analysis using a single $X$ type TDI channel. We performed runs on each of
the $X$, $Y$ and $Z$ channels to see how the particular noise and signal realization in each channel
impacted our ability to detect a stochastic background signal. The results of this study are shown
in Figure 8. We see that the detection threshold for a single channel is roughly a factor of
two worse than when all links are operational. This result appears to contradict the usual statement
that LISA can only detect stochastic backgrounds when the null channel is available, but it should
be remembered that we are using the very strong assumption that the instrument noise is stationary
and Gaussian, and the individual noise sources have known spectral shape. With these assumptions
we find that the high frequency portion of the spectrum fixes the shot noise levels to very high precision,
so any deviations in the total spectrum in the $1 \rightarrow 10$ mHz range caused by a scale invariant
stochastic background stand out in stark relief.

In future studies it would be interesting to see how the detection limits are affected by relaxing the
assumptions in our noise model. It would also be interesting to have a better sense of how well component
level engineering models, ground testing and in-flight commissioning studies can constrain the instrument
noise model.

\section{Null channel for Unequal Arm LISA}

\begin{figure}[htbp]
\label{unequalSens}
   \centering
   \includegraphics[width=2.3in,angle=270] {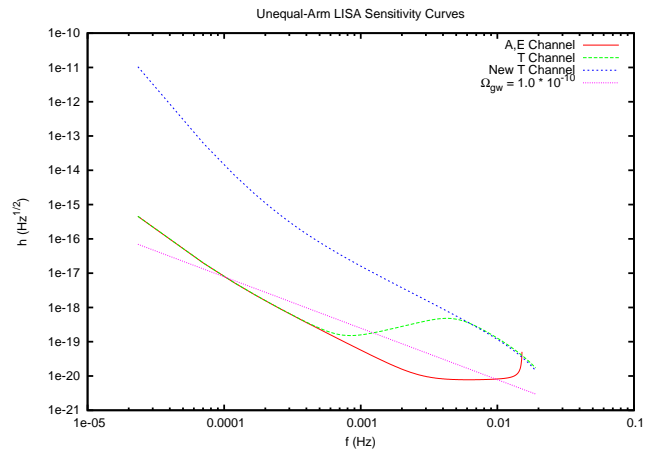} 
   \caption{Sensitivity curve for unequal arm LISA.  Our new $T$ channel restores the usual
low-frequency insensitivity to a stochastic gravitational wave background.}
\end{figure}

The various null channels that have been identified for LISA - the symmetric Sagnac channel;
the Sagnac $T$ channel and the Michelson $T$ channel - are only null if the arm-lengths of the
detector are equal. The orbits of the LISA spacecraft cause the arm-lengths to vary by a few
percent over the course of a year. In practice the arm-lengths will never be equal.
For unequal arm-lengths we find that the Michelson $T$ channel has {\em exactly}
the same sensitivity curve as the $A$ or $E$ channels, at least for low frequencies. We found that
it is possible to restore the relative insensitivity of the $T$ channel by forming a new, time
delayed combination of the $X$, $Y$ and $Z$ channels. Since the arm-lengths will be approximately
equal we write
\begin{eqnarray}
L_a    &=& L(1 + \epsilon_a)  \nonumber \\
L_b    &=& L(1 + \epsilon_b) \nonumber \\
L_c    &=& L(1 + \epsilon_c) \, ,
\end{eqnarray}
with $\vert \epsilon_i \vert \ll 1$. Here an ``a" subscript denotes the ``12" arm, ``b" the ``23" arm, and ``c" the ``13" arm.  
In the frequency domain we define a modified
$T$ channel:
\begin{equation}
T = \frac{1}{3}\left(X+\alpha Y+ \beta Z \right) 
\end{equation}
Working to leading order in $\epsilon_i$ and expanding $\alpha$ and $\beta$ in a Taylor series in $f/f_*$,
we are able to set the response function $R_{TT}$ to zero out to order $f^8$ (the
same as in the equal arm-length limit) with the coefficients
\begin{eqnarray}
\alpha &=& 1 + \epsilon\left[j_0 + ij_1\left(\frac{f}{f_*}\right) + j_2\left(\frac{f}{f_*}\right)^2 + j_4\left(\frac{f}{f_*}\right)^4 \right]\nonumber \\
\beta  &=& 1 + \epsilon\left[k_0 + k_2\left(\frac{f}{f_*}\right)^2 \right]
\end{eqnarray}
where $j_0$,$j_1$,$j_2$,$j_4$,$k_0$, and $k_2$ are given by:
\begin{eqnarray}
j_0 &=& \epsilon_a-\epsilon_b \nonumber \\
j_1 &=& \sqrt{\frac{1}{2}(\epsilon_a-\epsilon_b)^2+\frac{1}{2}(\epsilon_b-\epsilon_c)^2+\frac{1}{2}(\epsilon_a-\epsilon_c)^2} \nonumber \\
j_2 &=& \frac{1}{3}(\epsilon_b-\epsilon_a) \nonumber \\
j_4 &=& \frac{17}{3780}(\epsilon_b+\epsilon_c-2\epsilon_a) \nonumber \\
k_0 &=& \epsilon_c-\epsilon_b \nonumber \\
k_2 &=& \frac{1}{3}(\epsilon_b-\epsilon_c) \, .  \\
\end{eqnarray}
The $\alpha$ and $\beta$ coefficients define delay operators in the time domain.

Figure 9 shows the sensitivity curves for unequal arm LISA.  We see that at low frequencies,
the original $T$ channel has identical sensitivity to the $A$ and $E$ channels, while the new $T$ channel
restores the usual low-frequency insensitivity to a stochastic gravitational wave background.

\section{Next Steps}

\begin{figure}[t]
\begin{center}
\includegraphics[width=3.2in, height=2.8in]{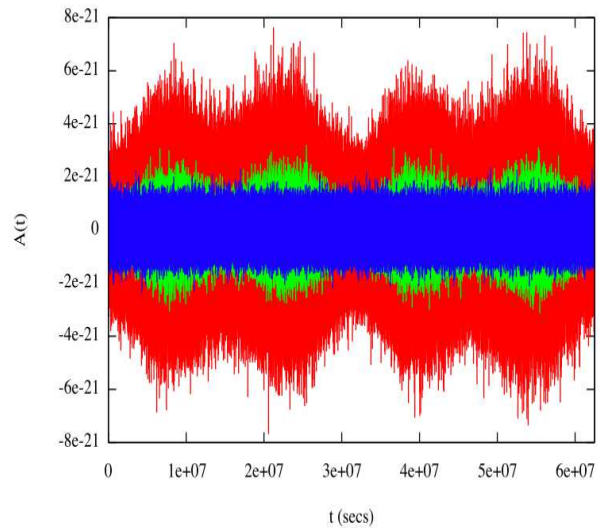}
\end{center}
\caption{The time-domain $A$-channel response to the full galactic foreground (red); the unresolved component of the
galactic foreground (green) and the instrument noise (blue). The signals were passband filtered between $0.1-10$ mHz.}
\label{fig:stochastic}
\end{figure}

We have shown that with a well constrained instrument noise model, and absent astrophysical foregrounds,
LISA could detect a scale invariant stochastic gravitational wave background as low as
$\Omega_{gw} = 6 \times 10^{-13}$ with just one month of data. One line of future study is to
investigate how this limit is affected by weakening the assumptions that went into the instrument noise
model. Another line of future study is to see how the limit is affected by astrophysical foregrounds,
such as the confusion noise from un-resolved white dwarf binaries in our galaxy.

It has been shown that $\sim 20,000$ of the brighter galactic binaries can be individually identified and
regressed from the LISA data~\cite{Crowder:2006eu}, but this still leaves behind a significant un-resolved
foreground signal. It should be possible to separate this signal from an isotropic cosmological stochastic
background by using the time-modulated response to the anisotropic galactic foreground. Figure~\ref{fig:stochastic}
shows the time domain response to the galactic foreground components. Notice that at certain times of the
year, the galactic confusion noise dips below the instrument noise for almost a full month. One
approach would be to restrict the analysis to these quiet periods, and achieve sensitives comparable to
what we found with a one-month data set. A more sophisticated approach is to generalize our model of
the cross-spectra to include a time-frequency representation of the response to the galactic foreground.
The parameters for this model may include the bulge and disk scale heights and radii, and the number density
of un-resolved signals as a function of frequency. Information from the resolved component could be used to
place priors on this model. Work on this problem is currently underway, and simulated data sets with a full
galactic foreground and an isotropic stochastic background have been generated for the fourth round
of the Mock LISA Data Challenge\cite{Babak:2009cj}.

Estimates of the extra-galactic white dwarf confusion noise~\cite{Farmer:2003pa} predict
$\Omega_{gw}(f) \sim 3 \times 10^{-12}$ at $f\sim 2-3{\rm mHz}$, which should be detectable
even with the galactic foreground to contend with. The extra-galactic astrophysical foreground
will likely set the floor for detecting stochastic backgrounds of cosmological origin, preventing
us from reaching the limits that the LISA observatory is theoretically capable of.

\section{Acknowledgments}
This work was supported by NASA grant NNX07AJ61G.

\bibliographystyle{apsrev}
\bibliography{GWBbib}

\appendix*
\section{Cross Spectra}

The noise cross-spectra are given by:

\begin{widetext}

\begin{equation}
 AA^*_p = \frac{4}{9} \sin^2\left(\frac{f}{f_*}\right)\bigg\{\cos\left(\frac{f}{f_*}\right)\bigg[4\big(S^p_{21}+S^p_{12}+S^p_{13}+S^p_{31}\big)
-2\big(S^p_{23}+S^p_{32}\big)\bigg]
         +5\big(S^p_{21}+S^p_{12}+S^p_{13}+S^p_{31}\big) +2\big(S^p_{23}+S^p_{32}\big)\bigg\}  
\end{equation}
		
\begin{eqnarray}
AA^*_a &=& \frac{16}{9}\sin^2\left(\frac{f}{f_*}\right)\Bigg\{\cos\left(\frac{f}{f_*}\right) \bigg[4\big(S^a_{12}+S^a_{13}+S^a_{31}+S^a_{21}\big)-2\big(S^a_{23}+S^a_{32}\big)\bigg] \nonumber \\       
       & & +\cos(2u)\bigg[\frac{3}{2}(S^a_{12}+S^a_{13}+S^a_{23}+S^a_{32})+2(S^a_{31}+S^a_{21})\bigg] \nonumber \\
			 & & +\frac{9}{2}(S^a_{12}+S^a_{13})+3(S^a_{31}+S^a_{21})+\frac{3}{2}(S^a_{23}+S^a_{32})\Bigg\}
\end{eqnarray}
 
\begin{eqnarray}                          
EE^*_s &=& \frac{4}{3}\sin^2\left(\frac{f}{f_*}\right)\bigg\{S^p_{21}+S^p_{12}+S^p_{13}+S^p_{31}+\big(S^p_{23}+S^p_{32}\big)\big(2+2\cos\left(\frac{f}{f_*}\right)\big)\bigg\}
\end{eqnarray}

\begin{eqnarray}
EE^*_a &=& \frac{16}{3}\sin^2\left(\frac{f}{f_*}\right)\bigg\{ S^a_{23}+S^a_{32}+S^a_{21}+S^a_{31} +2\cos\left(\frac{f}{f_*}\right)\bigg(S^a_{23}+S^a_{32}\bigg) \\ \nonumber
			 & & +\cos^2\left(\frac{f}{f_*}\right)\bigg(S^a_{23}+S^a_{32}+S^a_{12}+S^a_{13}\bigg)\bigg\}
\end{eqnarray}
	 
\begin{eqnarray}
TT^*_p &=& \frac{4}{9}\sin^2\left(\frac{f}{f_*}\right)\bigg(2-2\cos\left(\frac{f}{f_*}\right)\bigg)\bigg(S^p_{13}+S^p_{32}+S^p_{21}+S^p_{12}+S^p_{31}+S^p_{23}\bigg)
\end{eqnarray}

\begin{eqnarray}
TT^*_a &=& \frac{16}{9}\sin^2\left(\frac{f}{f_*}\right)\bigg(1-2\cos\left(\frac{f}{f_*}\right)+\cos^2\left(\frac{f}{f_*}\right)\bigg) \bigg(S^a_{12}+S^a_{13}+S^a_{31}+S^a_{32}+S^a_{23}+S^a_{21}\bigg)
\end{eqnarray}

\begin{eqnarray}
AE^*_p &=& -\frac{4}{3\sqrt{3}}\sin^2\left(\frac{f}{f_*}\right) \bigg(2\cos\left(\frac{f}{f_*}\right)+1\bigg)\bigg(S^p_{13}-S^p_{12}+S^p_{31}-S^p_{21}\bigg)
\end{eqnarray}

\begin{eqnarray}
AE^*_a &=& \frac{16}{3\sqrt{3}}\sin^2\left(\frac{f}{f_*}\right)\bigg\{2\cos\left(\frac{f}{f_*}\right)\bigg(S^a_{13}-S^a_{12}+S^a_{31}-S^a_{21}\bigg)\nonumber \\
			 & & +\cos^2\left(\frac{f}{f_*}\right)\bigg(S^a_{13}-S^a_{12}+S^a_{23}-S^a_{32}\bigg) +S^a_{31}-S^a_{21}-S^a_{23}+S^a_{32}\bigg\}
\end{eqnarray}

\begin{eqnarray}
AT^*_p &=& \frac{4}{9}\sin^2\left(\frac{f}{f_*}\right) \bigg(1-\cos\left(\frac{f}{f_*}\right)\bigg)\bigg(S^p_{12}+S^p_{13}+S^p_{21}+S^p_{31} -2\big(S^p_{32}+S^p_{23}\big)\bigg)
\end{eqnarray}

\begin{eqnarray}
AT^*_a &=& \frac{16}{9}\sin^2\left(\frac{f}{f_*}\right)
       \bigg\{\frac{3}{2} \bigg(S^a_{12}+S^a_{13}-S^a_{23}-S^a_{32}\bigg)
       +\cos\left(\frac{f}{f_*}\right)\bigg[2\bigg(S^a_{23}+S^a_{32}\bigg)-\bigg(S^a_{12}+S^a_{13}\bigg)-\bigg(S^a_{31}+S^a_{21}\bigg)\bigg] \nonumber \\
       & & +\cos(2u)\bigg[\bigg(S^a_{31}+S^a_{21}\bigg)-\frac{3}{2}\bigg(S^a_{12}+S^a_{13}\bigg)-\frac{3}{2}\bigg(S^a_{23}+S^a_{32}\bigg)\bigg]\bigg\}
\end{eqnarray}
		
\begin{eqnarray}
ET^*_p &=& \frac{4}{3\sqrt{3}}\sin^2\left(\frac{f}{f_*}\right)\bigg(\cos\left(\frac{f}{f_*}\right)-1\bigg)\bigg(S^p_{21}-S^p_{31}+S^p_{12}-S^p_{13}\bigg)
\end{eqnarray}

\begin{eqnarray}
ET^*_a &=& \frac{16}{3\sqrt{3}}\sin^2\left(\frac{f}{f_*}\right) 
       \bigg\{\cos^2\left(\frac{f}{f_*}\right)\bigg(S^a_{32}-S^a_{23}+S^a_{12}-S^a_{13}\bigg) 
        +\cos\left(\frac{f}{f_*}\right)\bigg(S^a_{31}-S^a_{21}+S^a_{13}-S^a_{12}\bigg) \nonumber \\
       & & +S^a_{23}-S^a_{32}+S^a_{21}-S^a_{31}\bigg\}
\end{eqnarray}

\end{widetext}

\end{document}